\documentstyle[12pt]{article}
\begin{document}\newcommand{\nc}{\newcommand}
\nc{\beq}{\begin{equation}}
\nc{\eeq}{\end{equation}}
\nc{\bea}{\begin{eqnarray}}
\nc{\eea}{\end{eqnarray}}
\nc{\ba}{\begin{array}}
\nc{\ea}{\end{array}}
\nc{\nn}{\nonumber}
\nc{\bpi}{\begin{picture}}
\nc{\epi}{\end{picture}}
\nc{\scs}{\scriptstyle}
\nc{\sss}{\scriptscriptstyle}
\nc{\sst}{\scriptstyle}
\nc{\ts}{\textstyle}
\nc{\ds}{\displaystyle}
\nc{\sctn}[1]{\section{\hspace{-18pt}.\ #1}}
\nc{\subsctn}[1]{\subsection{\hspace{-16pt}.\ #1}}
\nc{\subsubsctn}[1]{\subsubsection{\hspace{-14pt}.\ #1}}

\nc{\al}{\alpha}
\nc{\be}{\beta}
\nc{\ga}{\gamma}
\nc{\Ga}{\Gamma}
\nc{\de}{\delta}
\nc{\De}{\Delta}
\nc{\ep}{\epsilon}
\nc{\ve}{\varepsilon}
\nc{\eb}{\bar{\eta}}
\nc{\et}{\eta}
\nc{\ka}{\kappa}
\nc{\la}{\lambda}
\nc{\La}{\Lambda}
\nc{\th}{\theta}
\nc{\Th}{\Theta}
\nc{\ze}{\zeta}
\nc{\p}{\partial}

\nc{\cw}{c_\th}
\nc{\oocw}{\cw^{-1}}
\nc{\cwcw}{\cw^2}
\nc{\cwcwcw}{\cw^3}
\nc{\cwcwcwcw}{\cw^4}
\nc{\oocwcw}{\cw^{-2}}
\nc{\sw}{s_\th}
\nc{\oosw}{\sw^{-1}}
\nc{\swsw}{\sw^2}
\nc{\swswsw}{\sw^3}
\nc{\swswswsw}{\sw^4}
\nc{\tw}{t_\th}
\nc{\twtw}{\tw^2}

\nc{\mb}[1]{\mbox{#1}}
\nc{\unit}{\mb{\bf\large 1}}
\nc{\Tr}{\mb{Tr}}
\nc{\Bb}{\mb{\boldmath $\ds B$}}
\nc{\Wb}{\mb{\boldmath $\ds W$}}
\nc{\Xb}{\mb{\boldmath $\ds X$}}
\nc{\Fb}{\mb{\boldmath $\ds F$}}
\nc{\Tb}{\mb{\boldmath $\ds T$}}
\nc{\Vb}{\mb{\boldmath $\ds V$}}
\nc{\mw}{m_{\sss W0}}
\nc{\mz}{m_{\sss Z0}}
\nc{\half}{{\ts\frac{1}{2}}}
\nc{\ihalf}{{\ts\frac{i}{2}}}
\nc{\dg}{\dagger}
\nc{\Lag}{{\cal L}}
\nc{\Lgf}{\Lag_{\mbox{\scriptsize\em gf}}}
\nc{\Lgh}{\Lag_{\mbox{\scriptsize\em gh}}}
\nc{\Lgfgh}{\Lag_{\mbox{\scriptsize\em gf,gh}}}
\nc{\Lwhc}{\Lag_{\mbox{\scriptsize\em Whc}}}
\nc{\Lbhc}{\Lag_{\mbox{\scriptsize\em Bhc}}}
\nc{\Lvhc}{\Lag_{\mbox{\scriptsize\em Vhc}}}
\nc{\Luhc}{\Lag_{\mbox{\scriptsize\em Uhc}}}
\nc{\od}{{\cal O}}
\nc{\mubar}{\bar{\mu}}
\nc{\Laeff}{\La_{\rm eff}}

\nc{\ltap}{\;\raisebox{-.4ex}{\rlap{$\sim$}}\raisebox{.4ex}{$<$}\;}
\nc{\gtap}{\;\raisebox{-.4ex}{\rlap{$\sim$}}\raisebox{.4ex}{$>$}\;}

\begin{flushright}
Freiburg--THEP 99/11\\
UM-TH-99-06\\
September 1999
\end{flushright}

\vspace{1.5cm}

\begin {center}

{\Large \bf Considerations on anomalous vector boson couplings}\\

\vskip 1.cm
{\bf J.J. van der Bij}\\ 
\medskip
Albert--Ludwigs--Universit\"at Freiburg, 
Fakult\"at f\"ur Physik, \\
Hermann--Herder--Strasse 3, 79104 Freiburg i. Br., Germany\\
\vskip .7cm 
{\bf A. Ghinculov }\\
\medskip 
Randall Laboratory of Physics, University of Michigan,\\
Ann Arbor, Michigan 48109-1120, USA\\
\end{center}
\bigskip

\begin{abstract}
We discuss the meaning of anomalous vector boson self couplings.
Implications of present experimental constraints for 
future colliders are discussed. Results for triple vector boson production
at the LHC are given.
\end{abstract}
\medskip

\noindent{\large \bf 1. Introduction}\\

The standard model is well established by the experiments at LEP and the
Tevatron. Any deviations of the standard model can therefore be
introduced only with care. Changes to the standard model come with different
forms of severity. In order to see at what level anomalous vector boson 
couplings can be reasonably discussed one has to consider these cases
separately. Changes to the gauge structure of the theory, that do not
violate the renormalizability of the theory, i.e. the introduction of
extra fermions or possible extensions of the gauge group are the least severe.
They will typically generate small corrections to vector boson couplings
via loop effects. In this case also radiative effects will be generated
at lower energies. For the LHC the important thing in this case is not to 
measure the anomalous couplings precisely, but to look for the extra particles.
This subject belongs naturally to the "extensions of the standard model"
working group. We will not discuss it further. In the other case,
a more fundamental role is expected for the anomalous couplings, implying
strong interactions. In this case one has to ask oneself whether one should
study a model with or without a fundamental Higgs boson. 

Simply removing the
Higgs boson from the standard model is a relatively mild change. The model
becomes nonrenormalizable, but the radiative effects grow only logarithmically
with the cut-off. The question is whether this scenario is ruled out by the
LEP1 precision data. The LEP1 data appear to be in agreement with the standard
model, with a preferred low Higgs mass. One is sensitive to the Higgs mass
in three parameters , known as S,T,U or $\epsilon_1, \epsilon_2, \epsilon_3$.
These receive corrections of  the form $g^2 (log(m_H/m_W) + constant)$,
where the constants are of order one. The logarithmic enhancement is universal
and would also appear in models without a Higgs as $log(\Lambda)$, where
$\Lambda$ is the cut-off, where new interactions should appear. Only when 
one can determine the three different constants independently, can one say 
that one has established
the standard model. At present the data do not suffice to do this to great
enough precision.\\

A much more severe change to the standard model is the introduction of 
non-gauge vector
boson couplings. These new couplings violate renormalizability much more
severely than simply removing the Higgs boson. Typically quadratically
and quartically divergent corrections would appear to physical observables.
It is therefore questionable, if one should study models with a fundamental
Higgs boson, but with extra anomalous vector boson couplings. It is hard to
imagine a form of dynamics that could do this. If the vector bosons 
become strongly interacting the Higgs probably would at most exist in an
"effective" way. The most natural way is therefore to study anomalous
vector boson couplings in models without a fundamental Higgs. Actually
when one removes the Higgs boson the standard model becomes a gauged non-linear
sigma-model. The nonlinear sigma-model is well known to describe low-energy pion
physics. The "pions" correspond to the longitudinal degrees of freedom
of the vector bosons. To $f_{\pi}$ corresponds the vacuum expectation value
of the Higgs field. 
Within this description  the standard model corresponds to the lowest order term
quadratic in the momenta, anomalous couplings to higher derivative terms.
The systematic expansion in terms of momenta is known as chiral perturbation
theory and is extensively used in meson physics. 

Writing down the most general non-linear chiral Lagrangian containing
up to four derivatives gives rise to a large number of terms, which are too
general to be studied
effectively. One therefore has to look for dynamical principles
that can limit the number of terms. Of particular importance are approximate
symmetry principles. In the first place one expects CP-violation to be
small. We limit ourselves therefore to CP-preserving terms. 
In order to see what this means in practice it is
advantageous to describe the couplings in a manifestly gauge-invariant
way, using the St\"uckelberg formalism \cite{stuc}. One needs the
 following definitions:

\begin{equation} F_{\mu \nu}=\frac {i \tau_i}{2} (\partial_{\mu}W^i_{\nu} - \partial_{\nu}
W^i_{\mu} + g \epsilon^{ijk}W^j_{\mu} W^k_{\nu} ) \end{equation}
is the SU(2) field strength.
\begin{equation} D_{\mu} U = \partial_{\mu} U + \frac {ig}{2} \tau_i W^i_{\mu} U
 + ig tg \theta_w U \tau_3 B_{\mu} \end{equation}
Is the gauge covariant derivative of the SU(2) valued field $U$ , that
describes the longitudinal degrees of freedom of the vector fields
in a gauge invariant way.
\begin{equation} B_{\mu \nu} = \partial_{\mu} B_{\nu} - \partial_{\nu} B_{\mu} \end{equation}
is the hypercharge field strength.
\begin{equation} V_{\mu} = (D_{\mu} U) U^{\dagger} / g \end{equation}
\begin{equation} T = U \tau_3 U^{\dagger} / g \end{equation}
are auxiliary quantities having simple transformation properties.
Excluding CP violation, the nonstandard three and four vector boson couplings
are described in this formalism by the following set of operators.
\begin{equation}  {\cal L}_1  = Tr ( F_{\mu \nu} [V_{\mu},V_{\nu}]) \end{equation}
\begin{equation}  {\cal L}_2  = i\frac {B_{\mu \nu}}{2} Tr (T [V_{\mu},V_{\nu}]) \end{equation}
\begin{equation}  {\cal L}_3  = Tr ( T F_{\mu \nu} ) Tr (T [V_{\mu},V_{\nu}]) \end{equation}
\begin{equation}  {\cal L}_4  = ( Tr [V_{\mu} V_{\nu} ]) ^2 \end{equation}
\begin{equation}  {\cal L}_5  = ( Tr [V_{\mu} V_{\mu} ]) ^2  \end{equation}
\begin{equation}  {\cal L}_6  = Tr(V_{\mu} V_{\nu}) Tr(T V_{\mu}) Tr(T V_{\nu}) \end{equation}
\begin{equation}  {\cal L}_7  = Tr(V_{\mu} V_{\mu}) (Tr [T V_{\nu}]) ^2 \end{equation}
\begin{equation}  {\cal L}_8  = \frac {1}{2} [ (Tr [T V_{\mu}]) (Tr [T V_{\nu}])]^2 \end{equation}

In the unitary  gauge $U=1$, one has
\begin{equation} {\cal L}_1  = i [ ( c Z_{\mu \nu} + s F_{\mu \nu} ) W_{\mu}^+ 
  W_{\nu}^- + Z_{\nu}/c ( W_{\mu \nu}^+ W_{\mu}^- 
   - W_{\mu \nu}^- W_{\mu}^+) ] \end{equation}
~~~~~~~~~~~~~ + gauge induced four boson vertices 
\begin{equation} {\cal L}_2 = i ( c F_{\mu \nu} - 
         s Z_{\mu \nu}) W_{\mu}^+ W_{\nu}^-   \end{equation}
	 
\begin{equation} {\cal L}_3 = i ( c Z_{\mu \nu} + 
         s F_{\mu \nu}) W_{\mu}^+ W_{\nu}^-   \end{equation}	  
c and s are cosine and sine of the weak mixing angle. 
The standard model without Higgs corresponds to:
\beq
\label{lagew}
\ts\Lag_{EW}=
-\half\Tr(\Wb_{\mu\nu}\Wb^{\mu\nu})
-\half\Tr(\Bb_{\mu\nu}\Bb^{\mu\nu})
+\frac{g^2v^2}{4}\Tr(\Vb_\mu \Vb^\mu)
\eeq
\\

\noindent{\large \bf 2. Dynamical constraints}\\

The list contains terms that give rise to vertices with minimally three
or four vector bosons. Already with the present data a number of constraints
and/or consistency conditions can be put on the vertices. 
The most important of these come from the limits on the breaking of the
so-called custodial symmetry. If the hypercharge is put to zero the
effective Lagrangian has a larger symmetry than $SU_{L}(2) \times U_Y(1)$, i.e.
it has the symmetry $SU_{L}(2) \times SU_{R}(2)$. The $SU_{R}(2)$ invariance
is a global invariance. Within the standard model this invariance is an
invariance of the Higgs potential, but not of the full Lagrangian. 
It is ultimately this invariance that is responsible for the fact that the
$\rho$-parameter, which is the ratio of charged to neutral current strength,
is equal to one at the tree level. Some terms in the Lagrangian, i.e. the
ones containing the hypercharge field explicitly or the terms with $T$,
that project out the third isospin component violate this symmetry explicitly.
These terms, when inserted in a loop graph, give rise to quartically divergent
contributions to the $\rho$-parameter. Given the measurements this means that
the coefficients of these terms must be extremely small. It is therefore reasonable
to limit oneself to a Lagrangian, where hypercharge appears only indirectly
via a minimal coupling, so without explicit T. 
This assumption means physically, that the ultimate dynamics that is
responsible for the strong interactions among the vector bosons acts in the
non-Abelian sector. Indeed one would normally not expect precisely the
hypercharge to become strong. However we know, that there is a strong violation 
of the custodial symmetry in the form of the top-quark mass. Actually the 
top-mass almost saturates the existing corrections to the $\rho$-parameter,
leaving no room for violations of the custodial symmetry in the anomalous
vector boson couplings. We therefore conclude: {\it If there really are strong
vector boson interactions, the mechanism for mass generation is unlikely to be the
same for bosons and fermions}.

Eliminating the custodial symmetry violating interactions we are left with the
simplified Lagrangian, containing ${\cal L}_1$, ${\cal L}_4$, ${\cal L}_5$.
Besides the vertices there are in
principle also propagator corrections. We take the two-point functions
without explicit $T$.
Specifically, we add to the theory\cite{kastening1}
\beq
\label{laghctr}
\ts\Lag_{hc,tr}
=\frac{1}{2\La_W^2}\Tr[(D_\al\Wb_{\mu\nu})(D^\al\Wb^{\mu\nu})]
+\frac{1}{2\La_B^2}\Tr[(\p_\al\Bb_{\mu\nu})(\p^\al\Bb^{\mu\nu})]
\eeq
for the transverse degrees of freedom of the gauge fields and
\beq
\label{laghclg}
\ts\Lag_{hc,lg}
=-\frac{g^2v^2}{4\La_V^2}\Tr[(D^\al\Vb^\mu)(D_\al\Vb_\mu)]
\eeq
for the longitudinal ones, where the $\La_X$ parametrize
the quadratic divergences and are expected to represent the
scales where new physics comes in. In phenomenological applications
these contributions give rise to formfactors in the propagators
\cite{kastening1, vdbij}. Introducing such cut-off dependent
propagators in the analysis of the vector boson pair production is
similar to having s-dependent triple vector boson couplings,
which is the way the data are usually analysed. 

This effective Lagrangian is very similar to the one in pion-physics.
Indeed if one takes the limit, vev fixed and gauge couplings to zero,
one finds the standard pion Lagrangian. As it stands one can use the LEP1 data
to put a limit on the terms in the two point vertices. Using a naive analysis
one finds \cite{kastening1} $1/\Lambda_B^2=0$. For the other two cut-offs
one has:\\

{\bf A. The case $\Lambda_V^2>0,\Lambda_W^2<0$}\\
$\Lambda_V>0.49 TeV, \Lambda_W>1.3 TeV$\\

{\bf B. The case $\Lambda_V^2<0,\Lambda_W^2>0$}\\
$\Lambda_V>0.74 TeV, \Lambda_W>1.5 TeV$ \\

This information is important for further limits at high energy colliders, 
as it tells us, how one has to cut off off-shell propagators. We notice
that the limits on the form factors are different for the transversal, longitudinal
and hypercharge formfactors. The precise limits are somewhat qualitative
and should be taken as such. However they show that effective cut-off
form factors should be taken around 500 GeV. It is certainly not correct to
put them at the maximum machine energy. Further information comes from the direct measurements
of the three-point couplings at LEP2, which tell us that they are small. 
Similar limits at the Tevatron have to be taken with some care, as there is 
a cut-off dependence. As there is no known model that can give large three-point
interactions, we asume for the further analysis of the fourpoint vertices,
that the three point anomalous couplings are absent. On the remaining two 
fourpoint vertices two more constraints can be put. The first comes from 
consistency of chiral perturbation theory \cite{pelaez}. Not every effective chiral
Lagrangian can be generated from a physical underlying theory.

A second condition comes from the $\rho$-parameter. Even the existing violation
of the custodial symmetry, though indirect via the minimal coupling to hypercharge,
gives a contribution to the $\rho$-parameter. It constrains the 
combination $5g_4 + 2g_5$. The remaining combination $2 {\cal L}_4 - 5 {\cal L}_5$ 
is fully unconstrained by experiment and gives in principle a possibility
for very strong interactions to be present. However this particular combination does
not seem to have any natural interpretation from underlying dynamics. Therefore
one can presumably conclude that both couplings $ g_4,g_5$ are small.
There is a loophole to this conclusion, namely when the anomalous couplings
are so large that the one-loop approximation, used to arrive at the limits,
is not consistent and resummation has to be performed everywhere. This is a
somewhat exotic possibility, that could lead to very low-lying resonances,
which ought to be easy to discover at the LHC\cite{kastening2}.\\

\noindent{\large \bf 3. LHC processes}\\

Given the situation described above one  has to ask oneself, what the
LHC can do and in which way the data should be analysed. There are essentially
three processes that can be used to study vector bosonvertices:
vector boson pair production, vector boson scattering, triple vector boson production.
About the first two we have only a few remarks to make. 
They are discussed more fully
in other contributions to the workshop.\\

\noindent{\bf 3a. vector boson pair production}

Vector boson pair production can be studied in a relatively straightforward
way. The reason is that here the Higgs boson does not play a role in the
standard model, as we take the incoming quarks to be massless. Therefore
naive violations of unitarity can be compensated by the introduction of
smooth form-factors.

 One produces two vector bosons via normal standard model processes
with an anomalous vertex added. The extra anomalous coupling leads to
unitarity-violating cross-sections at high energy. As a total energy
of 14 TEV is available this is in principle  a serious problem. 
It is cured by introducing a formfactor for the incoming off-shell
line connected to the anomalous vertex. Naively this leads to a
form-factor dependent limit on the anomalous coupling in question.
The LEP1 data gives a lower limit on the cut-off to be used inside
the propagator. When one wants an overall limit on the anomalous coupling
one should use this value. This is particularly
relevant for the Tevatron. Here one typically takes
a cut-off of 2 TeV. This might give too strict a limit, as the LEP1 data
indicate that the cut-off can be as low as 500GeV. For practical
purposes the analysis at the Tevatron should give limits on anomalous
couplings for different values of the cut-off form factors, including
low values of the cut-off. For the analysis at the LHC one has much larger 
statistics. This means, that one can do better and measure limits on the
anomalous couplings as a function of the invariant mass of the produced system.
This way one measures the anomalous formfactor completely.\\ 
\noindent{\bf 3b. vector boson scattering}

Here the situation is more complicated than in vector boson pair production.
The reason is that within the standard model the process cannot be 
considered without intermediate Higgs contribution. This would violate unitarity.
However the incoming vector bosons are basically on-shell and this allows the 
use of unitarization methods, as are commonly used in chiral perturbation theory
in pion physics. These methods tend to give rise to resonances in longitudinal
vector boson scattering. The precise details depend on the coupling constants.
The unitarization methods are not unique, but generically give rise
to large I=J=0 and/or I=J=1 cross-section enhancements. The literature
is quite extensive. A good introduction is \cite{hikasa}; a recent review
is \cite{dominici}. \\

\noindent{\bf 3c. Triple vector boson production}

In this case it is not clear how one should consistently approach an analysis 
of anomalous vector boson couplings. Within the standard model the presence
of the Higgs boson is essential in this channel. Leaving it out one has to study
the unitarization. This unitarization has to take place not only
on the two-to-two scattering subgraphs, as in vector boson scattering, but
also on the incoming off-shell vector boson, decaying into three
real ones. The analysis here becomes too arbitrary to derive very meaningful
results. One cannot confidently
calculate anything here without a fully known underlying model of new strong 
interactions. Also measurable cross sections tend to be small, so that the
triple vector boson production is best used as corroboration of results in
vector boson scattering. Deviations of standard model cross sections
could be seen, but the vector boson scattering would be needed for interpretation.

One therefore needs the standard model results. The total number of
vector boson triples is given in table 1. We used an integrated luminosity
of $100 fb^{-1}$ and an energy of $14 TeV$ throughout.

\newpage
\begin{center}
\begin{tabular}{|l|l|l|l|l|}\hline
$m_{Higgs}$  &200 &400  &600  &800    \\ \hline
$W^+W^-W^-$  &11675  &5084   &4780  &4800    \\ \hline
$W^+W^+W^-$  &20250  &9243   &8684  &8768    \\ \hline
$W^+W^-Z$    &20915  &11167   &10638  &10685    \\ \hline
$W^-ZZ$	    &2294  &1181   &1113  &1113    \\ \hline
$W^+ZZ$	    &4084  & 2243  &2108  &2165    \\ \hline
$ZZZ$	    &4883  & 1332  &1087  &1085    \\ \hline
\end{tabular}
\vskip 0.4cm
Table 1. Total number of events, no cuts, no branching ratios.
\end{center}

One sees from this table that a large part of the events comes from associated
Higgs production, when the Higgs is light. However for the study of
anomalous vector boson couplings, the heavier Higgs results are arguably
more relevant.
Not all the events can be used for the analysis. If we limit ourselves
to events, containing only electrons,  
muons and neutrinos, assuming just acceptance cuts we find table 2.

\begin{center}
\begin{tabular}{|l|l|l|l|l|}\hline
$m_{Higgs}$  &200 &400  &600  &800    \\ \hline
$W^+W^-W^-$  &68  & 28  & 25 & 25  \\ \hline
$W^+W^+W^-$  & 112 &49   &44  & 44   \\ \hline
$W^+W^-Z$    & 32 & 17 &15  & 15   \\ \hline
$W^-ZZ$	    &1.0  &0.51   &0.46  &0.45    \\ \hline
$W^+ZZ$	    &1.7  & 0.88  &0.79  & 0.79   \\ \hline
$ZZZ$	    &0.62  &0.18   &0.13  & 0.12   \\ \hline
\end{tabular}
\vskip 0.4cm
Table 2. Pure leptons, $\vert \eta \vert < 3$, $p_T > 20 GeV$, no cuts on
neutrinos.
\end{center}

We see that very little is left, in particular in the processes with at least
 two Z-bosons, where the events can be fully reconstructed. In order to see 
how sensitive we are to anomalous couplings we assumed a 4Z coupling
with a formfactor cut-off at 2TeV. We make here no correction
for efficiencies etc. Using the triple Z-boson production, assuming
no events are seen in $100 fb^{-1}$,
we find a limit $\vert g_4 + g_5 \vert < 0.09$ at the 95\% CL,
where $g_4$ and $g_5$ are the coefficients muliplying the operators
${\cal L}_4$ and ${\cal L}_5$.
This is to be compared with $-0.15 < 5 g_4 + 2 g_5 < 0.14$ \cite{valencia}
 or $-0.066 < (5 g_4 + 2 g_5) \Lambda^2(TeV) < 0.026$ \cite{kastening1,vdbij}.
So the sensitivity is not better than present indirect limits. Better limits
exist in vector boson scattering \cite{eboli} or at a linear collider
\cite{ghinculov}.

In the following tables we present numbers for observable cross sections
in different decay modes of the vector bosons. We used the following cuts.

$$\vert \eta \vert_{lepton} < 3$$
$$\vert p_T \vert_{lepton} > 20 GeV$$
$$\vert \eta \vert_{jet} < 2.5$$
$$\vert p_T \vert_{jet} > 40 GeV$$
$$\Delta R_{jet,lepton} > 0.3 $$
$$\Delta R_{jet,jet} > 0.5 $$
$$\vert p_T \vert_{2 \nu} > 50 GeV$$

States with more than two neutrinos are not very useful because of the
background from two vector boson production. We did not consider final
states containing $\tau$-leptons.

With the given cuts the total number of events to be expected is rather small,
in particular since we did not consider the reduction in events due to
experimental efficiencies, which may be relatively large, because
of the large number of particles in the final state. For the
processes containing jets in the final state, there will be large
backgrounds due to QCD processes. A final conclusion on the significance
of the triple vector boson production for constraining the four
vector boson couplings will need more work, involving detector Monte 
Carlo calculations. 

However it is probably fair to say from the above
results, that no very strong constraints will be found from this process
at the LHC, but it is useful as a cross-check with other processes.
It may provide complementary information if non-zero anomalous couplings are
found.\\ 

\begin{center}
\begin{tabular}{|l|l|l|l|l|l|}\hline
$m_{Higgs}$           &200  &300  &400  &500  &600  \\ \hline
$6 \ell$              &0.62 &0.29 &0.18 &0.14 &0.13    \\ \hline
$4 \ell, 2 \nu$       &5.1  &2.5  &1.5  &1.2  &1.1    \\ \hline
$4 \ell, 2 j$         &6.6  &3.8  &2.2  &1.7  &1.4    \\ \hline
$2 \ell, 2 j, 2 \nu$  & 34  &20   &12   &9.0  &7.7    \\ \hline
$2 \ell, 4 j$	      & 24  &19   &11   &7.6  &6.0    \\ \hline
$2 \nu, 4j$	      & 37  &34   &21   &15   &11    \\ \hline
$6 j$	              & 25  &31   &19   &12   &8.7    \\ \hline
\end{tabular}
\vskip 0.4cm
Table 3. $ZZZ$ production in different decay modes.
\end{center}

\begin{center}
\begin{tabular}{|l|l|l|l|l|l|}\hline
$m_{Higgs}$           &200  &300  &400  &500  &600   \\ \hline
$4 \ell, 2 \nu$       &31   &20   &17   &16   &15    \\ \hline
$3 \ell, 2 j, 1 \nu$  &51   &40   &31   &28   &26    \\ \hline
$2 \ell, 4 j$         &19   &22   &17   &14   &13    \\ \hline
$2 \nu, 4 j$          &63   &74   &60   &51   &48    \\ \hline
$2 \ell, 2 j, 2 \nu$  &102  &68   &54   &49   &48    \\ \hline
$1 \ell, 4 j, 1 \nu$  &262  &196  &140  &127  &127   \\ \hline
$6 j$	              &86   &104  &78   &62   &56    \\ \hline
\end{tabular}
\vskip 0.4cm
Table 4. $WWZ$ production in different decay modes.
\end{center}

\begin{center}
\begin{tabular}{|l|l|l|l|l|l|}\hline
$m_{Higgs}$           &200  &300  &400  &500  &600   \\ \hline
$5 \ell, 1 \nu$ &0.45&1.04&0.63&0.52&0.47        \\ 
          &0.80&1.69&1.08&0.91&0.81              \\ \hline
$3 \ell, 2 j, 1 \nu$  &3.37&6.89&5.36&4.18&3.73    \\ 
                   &5.9&11.5&9.3&7.4&6.5         \\ \hline
$1 \ell, 4 j, 1\nu$ &7.6&11.5&12.4&10.0&8.4            \\ 
                    &13.3&20.0&21.6&18&15            \\ \hline
$4 \ell, 2 j$  &0.29&1.0&0.54&0.38&0.32            \\ 
               &0.49&1.6&0.91&0.65&0.54             \\ \hline
$2 \ell, 2 j, 2 \nu$ &2.0&6.5&3.5&2.5&2.2     \\ 
                     &3.4&10.7&6.1&4.4&3.7      \\ \hline
$2 \ell, 4 j$    &2.5&7.4&5.4&3.6&2.9 \\ 
                  &4.7&9.5&9.5&6.9&5.6       \\ \hline
$4 j, 2\nu$ &8.9&27&18&12.6&10.4        \\ 
             &195.&54&38&28&23           \\ \hline
$6 j$	  &5.3&12.3&13.3&8.8&7.4                 \\ 
         &9.1&20.7&23&16&12.5                  \\ \hline
\end{tabular}
\vskip 0.4cm
Table 5. $ZZW^-$(upper) and $ZZW^+$(lower) production in
different decay modes.
\end{center}

\begin{center}
\begin{tabular}{|l|l|l|l|l|l|}\hline
$m_{Higgs}$           &200  &300  &400  &500  &600   \\ \hline
$3 \ell, 3 \nu$       &66&44&37&35&33                \\  \hline
$\ell^+ \ell^+, 2 j, 2 \nu$ &57&43&31&26&24          \\ \hline
$\ell^+ \ell^-, 2 j, 2\nu$   &13&7.9&5.3&4.4&4.0     \\ \hline
$\ell^+, 4 j, 1 \nu$         &148&129&86&66&58   \\ \hline
$\ell^-, 4 j, 1 \nu$    &99&61&36&26&23  \\ \hline
$6 j$	               &50&74&46&32&25   \\ \hline
\end{tabular}
\vskip 0.4cm
Table 6. $W^-W^+W^+$ production in different decay modes.
\end{center}

\begin{center}
\begin{tabular}{|l|l|l|l|l|l|}\hline
$m_{Higgs}$           &200  &300  &400  &500  &600   \\ \hline
$3 \ell, 3 \nu$      &40&26&22&21&20                \\  \hline
$\ell^- \ell^-, 2 j, 2 \nu$ &34&25&17&14&13          \\ \hline
$\ell^+ \ell^-, 2 j, 2\nu$  &78&45&30&25&23      \\ \hline
$\ell^-, 4 j, 1 \nu$       &90&76&49&37&33   \\ \hline
$\ell^+, 4 j, 1 \nu$   &59&35&20&15&13   \\ \hline
$6 j$	               &29&43&26&18&14  \\ \hline
\end{tabular}
\vskip 0.4cm
Table 7. $W^+W^-W^-$ production in different decay modes.
\end{center}

\end{document}